\documentclass[12pt]{iopart}

\usepackage{graphicx}
\begin{document}

\title{Entanglement entropy scaling of the XXZ chain}
\author{Pochung Chen$^{1,2,5}$, Zhi-long Xue$^{1,2}$, I. P. McCulloch$^3$,  Ming-Chiang Chung$^{4,5} $, Miguel Cazalilla$^1$,  S.-K. Yip $^6$}
\address{$^1$ Department of Physics, National Tsing Hua University, Hsinchu 30013, Taiwan}
\address{$^2$ Frontier Research Center on Fundamental and Applied Sciences of Matters, National Tsing Hua University, Hsinchu 30013, Taiwan}
\address{$^3$ Centre for Engineered Quantum Systems, School of Mathematics and Physics, The University of Queensland, St Lucia, Queensland 4072, Australia}
\address{$^4$ Physics Department, Chung-Hsing University, Taichung,  40227, Taiwan}
\address{$^5$ Physics Division, National Center for Theoretical Sciences, Hsinchu 30013, Taiwan}
\address{$^6$ Institute of Physics and Institute of Atomic and Molecular Sciences, Academia Sinica, Taipei 11529, Taiwan}

\ead{pcchen@phys.nthu.edu.tw}

\begin{abstract}
We study the entanglement entropy scaling of the XXZ chain.
While in the critical XY phase of the XXZ chain the entanglement entropy scales logarithmically with a coefficient
that is determined by the associated conformal field theory, at the ferromagnetic point, however,
the system is not conformally invariant yet the entanglement entropy still scales logarithmically 
albeit with a different coefficient.
We investigate how such an nontrivial scaling at the ferromagnetic point influences the estimation of the central charge $c$
in the critical XY phase. In particular we use the entanglement scaling of the finite or infinite system, as well as the
finite-size scaling of the ground state energy to estimate the value of $c$. In addition, the spin-wave velocity and
the scaling dimension are also estimated. We show that in all methods the evaluations are influenced by the nearby
ferromagnetic point and result in crossover behavior. Finally we discuss how to determine whether the central charge estimation
is strongly influenced by the crossover behavior and how to properly evaluate the central charge.
\end{abstract}

\maketitle

\section{Introduction}

Entanglement plays an important role in distinguishing the nature of
quantum versus classical systems. It is an essential ingredient for quantum
computation. It also connects quantum information theory to the
traditional quantum many-body systems, for example, quantum critical phenomena
\cite{Osterloh:2002gs,Osborne:2002fs,Vidal:2003cn,Barthel:2006hz}
and topological systems such as fractional Quantum Hall effects \cite{Samuelsson:2004is,Kim:2004bi},
topological insulators \cite{Qi:2012jv} and graphene \cite{Ryu:2006bj, Chung:2011kq}.
In recent developments entanglement has also been related to numerical methods
based on tensor network algorithms, ranging from the density-matrix renormalization group
and matrix product state\cite{Schollwock:2011gl}, to the projected entangled pair states, and variational renormalization group methods \cite{Verstraete:2008ko}.

One can measure the entanglement of a pure state using bipartite
entanglement entropy:  Consider a pure state $|\Psi_{AB} \rangle$ of a bipartite total system $AB$ that consists of the
system $A$ and the environment $B$. The reduced density matrix of the
system A is $\rho_A = \Tr_B |\Psi_{AB} \rangle\langle \Psi_{AB} |
$. The entanglement entropy  $S_A \equiv   -\Tr \rho_A \log_2
\rho_A$, has been widely used to measure the bipartite entanglement
between the system $A$ and the environment $B$.
The entanglement entropy is especially useful for studying quantum
criticality. In one dimension, scaling of entanglement entropy is well
understood both for fermions and for bosons. For one-dimensional
quantum chains at zero temperature, it is generally known that  the
entanglement entropy $S_A$  of a system $A$  saturates  away from criticality, however, it scales
logarithmically when the system becomes quantum-critical, that is, when the
correlation length diverges. In the latter case, conformal field (CFT)
theory  \cite{Calabrese:2004hl} yields
\begin{equation}
   S_A(l)  = \frac{c+\bar{c}}{6} \log l + k,\label{eqn:s_scalinginfinite}
\end{equation}
where $c$ and $\bar{c}$ are holomorphic and antiholomorphic central
charges of the CFT and $k$ is a model-dependent constant.
For a quantum critical system with an unknown central charge, one can
estimate $c$ by calculating the scaling of entanglement entropy. For
the conveniency of numerical calculations, the scaling law of
entanglement entropy  for finite sizes \cite{Calabrese:2004hl}
\begin{equation}
  S_A(l,L)=\frac{c+\bar{c} }{6} \log\left[ \frac{L}{\pi}
    \sin\left(\frac{\pi l}{L} \right)  \right]+ k,
 \label{eqn:s_scalingfinite}
\end{equation}
for a subsystem of size $l$ and 
 total size $L$ (the size of the system plus environment)   is often used instead of the infinite size one
Eq.~(\ref{eqn:s_scalinginfinite}). In our system discussed below
$c=\bar{c}$.

However, the general knowledge about the scaling law of entanglement entry for
non-critical systems is not 100 percent correct. There are some
exceptions.   It is pointed out recently that even when the system loses the conformal invariance,
the entanglement entropy $S_A$ can still scale logarithmically with
the system size.  Examples include the ferromagnetic Heisenberg model
\cite{Popkov:2005dj,CastroAlvaredo:2011bb,CastroAlvaredo:2012ge}
and the infinite random fixed point (IRFP) of several random spin
chain models \cite{Refael:2004hu,Santachiara:2006bs, Fidkowski:2008hy}.
Those systems are at the crossover points between critical and
gapped systems and, most essentially, their ground states are highly degenerate.
Therefore it is  possible to define  an
effective central charge  for those gapped but  symmetric  systems, according to Eq.~(\ref{eqn:s_scalinginfinite}),
as three times the coefficient in front of the logarithmic scaling.
In this work we are interested in the situation where the CFT
regime ends at a ferromagnetic point with an effective central charge
that is larger than the central charge of the CFT regime.
We found that, if one only uses Eq.~(\ref{eqn:s_scalingfinite}) to
calculate central charges for finite systems, the central charges are
influenced by the ferromagnetic points. Hence, we study the effects of such a nearby ferromagnetic point
on the estimation of the value of the central charge from the numerical simulation.
We find that, depending on the algorithms and the physical quantities used to extract the value of central charge,
different crossover behavior arises as the system approaches the ferromagnetic point from within the critical regime.
Based on these results we show how to determine  if the
central charge estimation is strongly influenced  by the crossover
behavior and how to properly evaluate the central charge.

We consider the 1D spin-1/2 anisotropic Heisenberg (XXZ) model with
periodic boundary conditions (PBC).
The Hamiltonian reads:
\begin{equation}
  H=\frac{1}{2} \sum_{i=1}^L \left( S^+_i S^-_{i+1} + S^-_i S^+_{i+1} \right) + \Delta \sum_{i=1}^L S^z_i S^z_{i+1},
\end{equation}
where $\Delta$ is the anisotropy.
For $\Delta>+1$ the system is in the N\'eel phase which spontaneously breaks the lattice translation symmetry
and the ground states are two-fold degenerate.
For $\Delta<-1$ the system is in the ferromagnetic Ising phase which spontaneously breaks the spin reflection symmetry.
The ground states are two-fold degenerate and  fully polarized in $\pm z$ directions.
When $-1<\Delta \le +1$ the system is in the gapless critical XY phase.
It is known that the critical XY phase is described by a $c=1$ CFT.
The point $\Delta=-1$ is very special and is the main interest of this work.
We will refer this point as the {\em ferromagnetic point} in the rest of the manuscript.
At $\Delta=-1$ the symmetry of the Hamiltonian is enlarged to isotropic ferromagnet with full rotational symmetry.
The ground states are infinitely degenerate in the thermodynamic limit.
However the system is not conformally invariant at this point.
When the system is in the ferromagnetic phase $(\Delta<-1)$ the entanglement saturates as one increases the block size.
However, it has been shown when $\Delta=-1$ there is an essential singularity in the
entanglement entropy \cite{Ercolessi:2011er, Ercolessi:2012bq}.
It has been also shown that in the limit of $\Delta \rightarrow -1^+$ the entanglement entropy
scales logarithmically with the block size but with a coefficient that
is larger than the critical regime \cite{Popkov:2005dj,CastroAlvaredo:2011bb}.
For $\Delta \in (-1,+1]$ the ground state is $U(1)$ symmetric and has $S^{tot}_z=0$. 
At the ferromagnetic point the symmetry of the Hamiltonian is enlarged to SU(2), 
but this is broken by the ground state and hence there are infinitely many degenerate 
ground states in the thermodynamic limit. It is exactly this ground stage degeneracy 
that gives rise to the logarithmic scaling of the entanglement entropy. 
In this work we are interested in the entanglement entropy scaling of the 
ground state that is smoothly connected to the ground state in the critical XY phase. 
This particular ground state at the ferromagnetic point can be reached by taking the 
$\Delta -> -1^+$ limit, which corresponds to the ferromagnetic ground state with $S_z^{tot}$ = 0.
It is predicted that as $\Delta \rightarrow -1^+$ one has
\begin{equation}
  \label{eq:scaling_fm}
  S_A(l) \sim \frac{1}{2} \log l.
\end{equation}
This corresponds to an effect central charge $c^F_{\mathrm{eff}}=3/2$ which is larger than
the $c=1$ in the XY phase. Consequently from entanglement entropy scaling point of view, there is a jump
from $c=1$ for $\Delta \in (-1,+1]$ to $c^F_{\mathrm{eff}}=3/2$ as $\Delta \rightarrow -1^+$.
For any numerical simulation, however, it is expected that such an abrupt jump is smeared out.
The resulting crossover behavior may depend on the algorithms and finite-size effects.
Consequently, conventional methods to extract the value of central charge
may be influenced by the nearby ferromagnetic point. To the best of
our knowledge,  such an influence is not widely studied in the literature.
Furthermore, it is pointed out in Ref.\cite{CastroAlvaredo:2012ge} that at $\Delta=-1$ it is possible to have
\begin{equation}
  \label{eq:scaling_fm_d}
  S_A(l) \sim \frac{d}{2} \log l,
\end{equation}
where $0\le d \le 2$. The exact value of $d$ depends on the particular ground state one choose
and $d$ can be interpreted as the (not necessarily an integer) number of zero-energy
Goldstone bosons describing the ground state. It is, however, not clear how such a prediction
manifest itself in the conventional calculation of the entanglement entropy.

To investigate these issues, we use three different methods to evaluate the central charge
in the regime $\Delta \in (-1,-1/2]$ in this work.
The first two methods study the entanglement scaling of a finite and an infinite system respectively
while the third method studies the scaling of the ground state energy.
Two numerical algorithms are used to evaluate relevant quantities for finite systems and infinite systems.
They are the density matrix renormalization group (DMRG) algorithm \cite{Schollwock:2005fa} for finite systems
and iDMRG algorithm \cite{McCulloch:2008va} for the infinite systems respectively.
(Note that iDMRG is not the infinite size DMRG algorithm that is used in the warm up stage of the DMRG algorithm.)
We pay special attention to the behavior
when $\Delta\rightarrow -1^+$. For the rest of the manuscript we will express
this limit as $\Delta+1\rightarrow 0^+$ for clarity.
The manuscript is organized as follows:
In Sec.\ref{sec:ent_finite}, we use the entanglement entropy scaling of a finite-system
to estimate the value of the central charge.
In Sec.\ref{sec:ent_infinite}, we study the entanglement entropy scaling of an infinite-system.
In Sec.\ref{sec:ent_gs}, we use finite-size scaling of the ground state energy to extract the central charge value.
Spin-wave velocity and the scaling dimension of the primary field are also estimated in Sec.\ref{sec:scaling}.
In Sec.\ref{sec:sum} we discuss our results and suggest a strategy to
determine accurately the value of the central charge when there is a ferromagnetic point nearby.

\section{Entanglement entropy scaling of a finite-system}
\label{sec:ent_finite}

\begin{figure}[btp]
\begin{center}
\includegraphics[width=0.9\columnwidth]{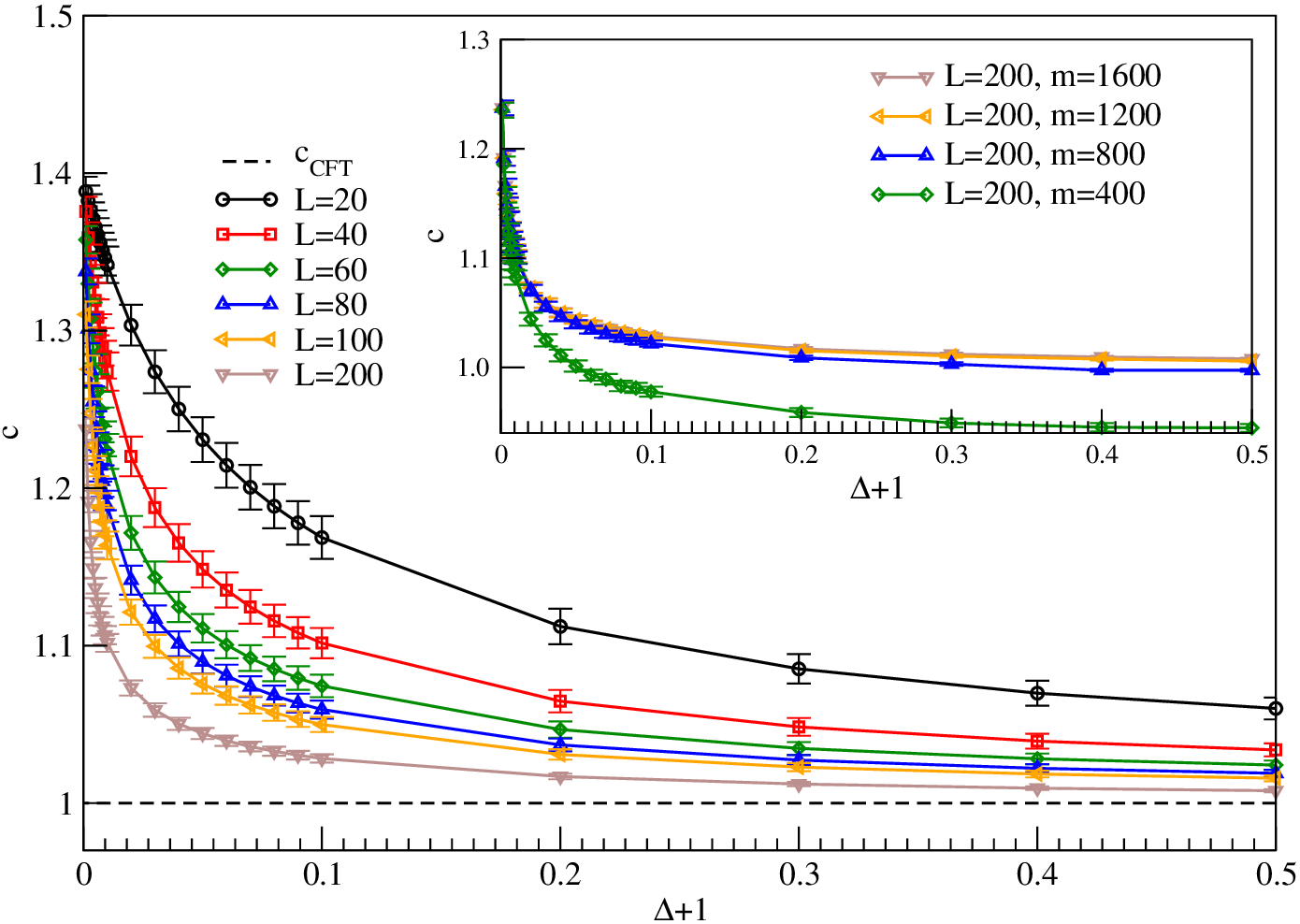}
\caption{(Color online)
$c_{\mathrm{DMRG}}(L)$ as a function of $\Delta+1$ with system size $L=20,40,60,80,100$, and $200$ respectively.
Inset: $c_{\mathrm{DMRG}}(L)$ with system size $L=200$, extracted from DMRG while keeping $m=400,800,1200$
states respectively.
}
\label{fig:c_eff}
\end{center}
\end{figure}

In this section we use the entanglement entropy scaling of a finite-size system to estimate the central charge $c$.
Most of the numerical studies in the literation to confirm the CFT predictions on XXZ chain concentrates on the regime of $\Delta \ge 0$. 
Some works extend the study to $\Delta <0$ \cite{Peschel:2005cc,Chiara:2006fo} but are not focused on the limit of $\Delta \rightarrow -1$,
which is the main interest of this work.
Here the DMRG algorithm \cite{Schollwock:2005fa} is used to obtain the ground state wave-function of a finite system with length $L$ with periodic boundary condition (PBC).
It is then straightforward to calculate the entanglement entropy $S_A(l,L)$ between a block of length $l$
and the rest of the system. 
In this work we use DMRG that preserves the $U(1)$ symmetry and target the $S^{tot}_z=0$ sector.
This is to ensure that the proper ground state is reached as we take the $\Delta \rightarrow -1^+$ limit.
We define an $L$-dependent effective central charge $c_{\mathrm{DMRG}}(L)$ by fitting the data using Eq.\ref{eqn:s_scalingfinite}.
The subscripts DMRG is used to distinguish from the central charge obtained by other methods.
Since the accuracy of the DMRG ground state depends on the number of states kept (denoted as $m$),
it is important to study how $c_{\mathrm{DMRG}}(L)$ depends on $m$.
For a given $L$, the entanglement entropy $S_A(l,L)$ reaches its maximum at half-chain $l=L/2$
and grows logarithmically with $L$ in the critical regime. On the other hand the maximal half-chain entanglement entropy
attainable by DMRG is $m \ln m$. It is then natural to expect that
for larger $L$ the value of $c_{\mathrm{DMRG}}(L)$ may depend strongly on the $m$ used.
In inset of Fig.~\ref{fig:c_eff} we show the value of $c_{\mathrm{DMRG}}(L=200)$ evaluated
using $m=400,800,1200$, and $1600$ respectively. We observe that for a fixed $\Delta$ the value of $c$
monotonically increases as $m$ increases. We find that while the data of $m=400$ seem
to fit Eq.\ref{eqn:s_scalingfinite} well (not shown here), the fitted $c_{\mathrm{DMRG}}$ seems to deviate substantially
from the results obtained with larger $m$. This indicates that $m=400$ is too small to
accurately determine the value of $c_{\mathrm{DMRG}}$. In contrast for $m\ge 800$ the
value starts to converge.  We also find that for smaller $L\le 80$, $m=400$ and $800$ lead to
similar values of $c_{\mathrm{DMRG}}$ and it is unnecessary to go to larger $m$.  These results suggest that
in order to estimate reliably the central charge, one should increase $m$ until the value of $c_{\mathrm{DMRG}}$
start to saturate or perform the extrapolation of $m$ to infinity.

In Fig.~\ref{fig:c_eff}, we plot $c_{\mathrm{DMRG}}(L)$ as a function of $\Delta+1$ for various $L$.
In this plot we ensure that sufficiently large $m$ is used for each $L$.
For a fixed $L$, we observe that $c_{\mathrm{DMRG}}(L)$ monotonically increases as $\Delta+1$ decreases
to zero from the positive side. Consequently the central charge deviates more and more from the CFT
prediction $c=1$ when approaching the ferromagnetic point.
Furthermore,  as $\Delta+1\rightarrow 0^+$, $c_{\mathrm{DMRG}}(L)$  seems to approachf $3/2$, the value
predicted in Ref.\cite{Popkov:2005dj,CastroAlvaredo:2011bb} regardless the $L$ used.
Recall that we use DMRG with U(1) symmetry and we target the ground state
with $S^{tot}_z=0$, which is exactly the state considered in Ref.\cite{Popkov:2005dj}. Hence at $\Delta=-1$ one expects
$S_A(l)\sim 1/2 \log l \sim c_{\mathrm{eff}}/3 \log l$ with $c_{\mathrm{eff}}=3/2$.
We also observe that for a fixed $\Delta$, $c_{\mathrm{DMRG}}(L)$ decreases monotonically as $L$ increases
and the crossover from $c=1$ to $c^F_{\mathrm{eff}}=3/2$ become sharper and shaper as $L$ increases.
Based on all the observations above,
we expect that in the thermodynamic limit the function $c_{\mathrm{DMRG}}$  becomes non-analytic at $\Delta=-1$,
resulting in $c=1$ for $\Delta \in (-1,1]$ and jump to $c=3/2$ at $\Delta=-1$.
This picture is consistent with the theoretical prediction of Ref.\cite{CastroAlvaredo:2011bb}.
This is also consistent with the calculation in Ref.\cite{Alba:2012jf}, in which by exact diagonalization of small $L$
systems it is found that the entanglement spectrum deviates substantially from the CFT
prediction when $\Delta<0$.
From the numerical  point of view, the important observation is that for smaller $L$ the crossover regime is larger and
one can overestimate the central charge by a large amount. It is also important to note that if the number of states kept
for the DMRG calculation is too small, one can underestimate the central charge.
Since the error due to finite $L$ and finite $m$ partially cancel each other,
a systematic study of $m$ and $L$ dependence of the central charge
is necessary to reliably extract the value of the central charge near a ferromagnetic point.

\section{Entanglement entropy scaling of an infinite-system}
\label{sec:ent_infinite}

\begin{figure}[btp]
\begin{center}
\includegraphics[width=0.9\columnwidth]{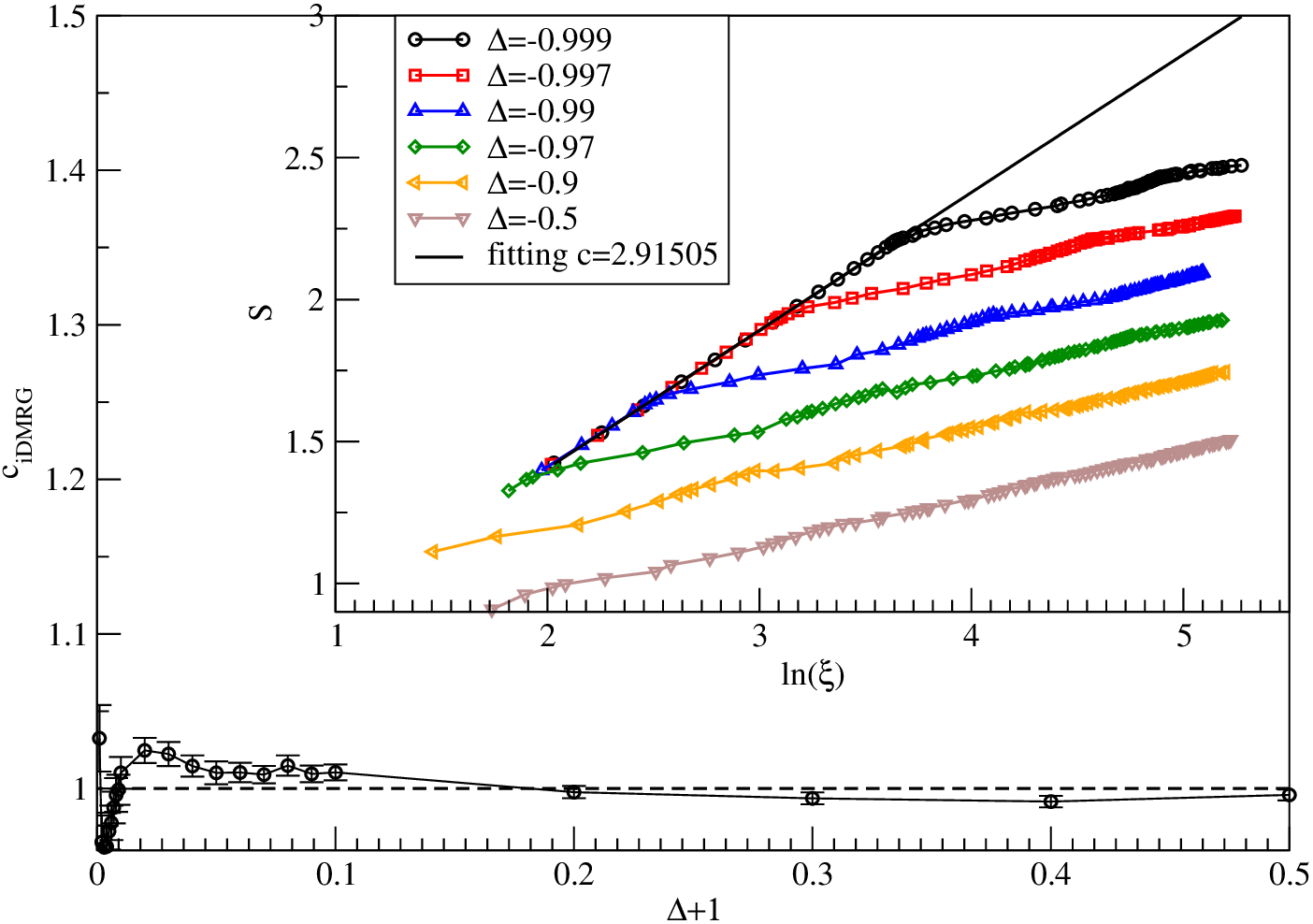}
\caption{(Color online)
$c_{\mathrm{iDMRG}}$ as a function of $\Delta+1$.
Inset: Entanglement entropy $S$ of a half-infinite chain  as a function of $\ln(\xi)$ for various $\Delta$.
Here $\xi$ is the correlation the length.
Black solid line is obtained by fitting data in the regime of  $\xi < \xi^F_c(\Delta)$
with $\Delta=0.999$. Here $\xi^F_c(\Delta)$ is a $\Delta$-dependent length scale as described in the main text.
}
\label{fig:iDMRG_all}
\end{center}
\end{figure}

In this section we use the entanglement entropy of an infinite system to estimate the value of central charge
in the thermodynamic limit. We employ the iDMRG algorithm to obtain the optimal ground state wavefunction
of an infinite system \cite{McCulloch:2008va}. The wavefunction is in the form of the matrix product state (MPS)
with truncation dimension $\chi$. 
If the true ground state is critical, the ground state obtained by iDMRG corresponds to a nearly critical system
with a large correlation length $\xi$. 
The ground state obtained by iDMRG can be imagine as the ground state of a nearly critical Hamiltonian, 
which is obtained by adding certain relevant perturbation to the critical Hamiltonian.
To ensure that the ground state is smoothly connected to the ground state in the critical XY phase,
we always preserve $U(1)$ symmetry in the iDMRG calculation and target the $S^{tot}_z=0$ sector.
For such a system the entanglement entropy of a half-infinite chain scales as \cite{Calabrese:2008cm}
\begin{equation}
  S=\frac{c}{6}\ln\xi.
  \label{eq:iDMRG_xi}
\end{equation}
In iDMRG both the half-infinite chain entropy $S$ and the correlation length $\xi$ can be easily calculated
from the transfer matrix. Eq.\ref{eq:iDMRG_xi} then can be used to estimate the central charge $c$.
We note that for such an infinite size algorithm, it is believed that the simulation at infinite $\xi$ and finite $\chi$ reproduces
the results at finite $\xi$ and infinite $\chi$ with a scaling law $\xi \propto \chi^k$.
Is is shown in Ref~\cite{Pollmann:2009ht} that
\begin{equation}
 k=\frac{6/c}{\sqrt{12/c+1}},
\end{equation}
consequently one has
\begin{equation}
  S=\frac{ck}{6}\ln\chi=\frac{1}{\sqrt{12/c}+1}\ln\chi,
\end{equation}
which can also be used to evaluate the value of central charge without evaluating the correlation length.
We find that fitting using Eq.\ref{eq:iDMRG_xi}  converges faster, but the results are always consistent with each other.

In the inset of Fig.~\ref{fig:iDMRG_all} we show the half-infinite chain entanglement entropy $S$
as a function of $\ln(\xi)$ for various $\Delta \in (-1,-1/2]$. We observe an interesting phenomenon:
When the system is far from the ferromagnetic point (for example $\Delta=-0.5$)
all the data fall on a straight-line. By fitting the data with Eq.~\ref{eq:iDMRG_xi}
one finds $c_{\mathrm{iDMRG}} \approx 1$ as predicted by the CFT.
When the system gets closer to the ferromagnetic point, however, the data starts to spilt into two regions which are separated by
a $\Delta$-dependent crossover length scale $\xi^F_c(\Delta)$.
In both regions the data fall on a straight line but with different slopes.
In Fig.~\ref{fig:iDMRG_all} we plot $c_{\mathrm{iDMRG}}$ which is obtained by fitting with Eq.\ref{eq:iDMRG_xi} but
using only data with $\xi>\xi^F_c(\Delta)$ as a function of $\Delta+1$. We always find $c_{\mathrm{iDMRG}}\approx 1$,
but larger and larger $\chi$ is needed to access the regime of $\xi>\xi^F_c(\Delta)$ as the system gets closer to the ferromagnetic point.
We conclude that $c_{\mathrm{iDMRG}} \approx 1$ provided that only data with $\xi>\xi^F_c(\Delta)$ are used for the fitting.
This is consistent with the CFT prediction and the results in the proceeding section
(after taking the limit of $L\rightarrow \infty$). We find that when the system gets very
close to the ferromagnetic point, the deviation from the expected result $c=1$ becomes larger.
This is due to (1) larger $\chi$ is needed to have enough data with $\xi>\xi^F_c(\Delta)$
and (2) there are many nearly degenerate low energy states when the system approaches
the ferromagnetic point, making it more difficult for iDMRG to converge.

Furthermore, it is surprising to observe that  all the data with $\xi<\xi^F_c(\Delta)$
seem to fall on a universal straight line regardless the value of $\Delta$.
Since the crossover length scale $\xi^F_c(\Delta)$ diverges as $\Delta \rightarrow -1^+$
we expect that the scaling behavior of the ferromagnetic point can be obtained
by fitting the data with $\xi<\xi^F_c(\Delta)$. We find that
\begin{equation}
  S(\Delta=-1)=\frac{2.915}{6}\ln\xi
\end{equation}
leading to an effective central charge $c^F_{\mathrm{eff}} = 2.915$. 
To understand this result we note that U(1) symmetry is preserved in our iDMRG calculation.
Similar to the finite-size DMRG we also target the $S^{tot}_z=0$ ground state.
We conjecture that for infinite size XXZ chains near the ferromagnetic point, 
when $\xi<\xi^F_c(\Delta)$ the entanglement scales as
\begin{equation}
  S \sim \frac{1}{2}\log \xi.
  \label{eq:scaling_fm_xi}
\end{equation}
If this scaling is interpreted as $S=\frac{c}{6}\log \xi$, one find $c^F_{\mathrm{eff}} = 3$
which is very close to the numerical value we obtain.
Our results also agree with the physical picture proposed in Ref~\cite{CastroAlvaredo:2011bb},
where it is proposed that when $\Delta \rightarrow -1^+$, there is an increasing characteristic length scale such that
(1) on the scales much below it, sites are randomly up and down with appropriate coefficients that locally reproduce
the ground state of $\Delta=-1$ and (2) on the scale above, a spin singlet that is expected for the ground state of the
XXZ model for $\Delta \in (-1,1]$. We believe that this characteristic length scale is exactly the $\xi^F_c(\Delta)$
we identified by iMDRG.  Consequently when $\xi<\xi^F_c(\Delta)$ the entanglement 
should scale according as Eq.\ref{eq:scaling_fm_xi} with an effective central charge $c^F_{\mathrm{eff}} = 3$ as observed in iDMRG calculation.



\section{Finite size scaling of the ground state energy}
\label{sec:ent_gs}

\begin{figure}[btp]
\begin{center}
\includegraphics[width=0.9\columnwidth]{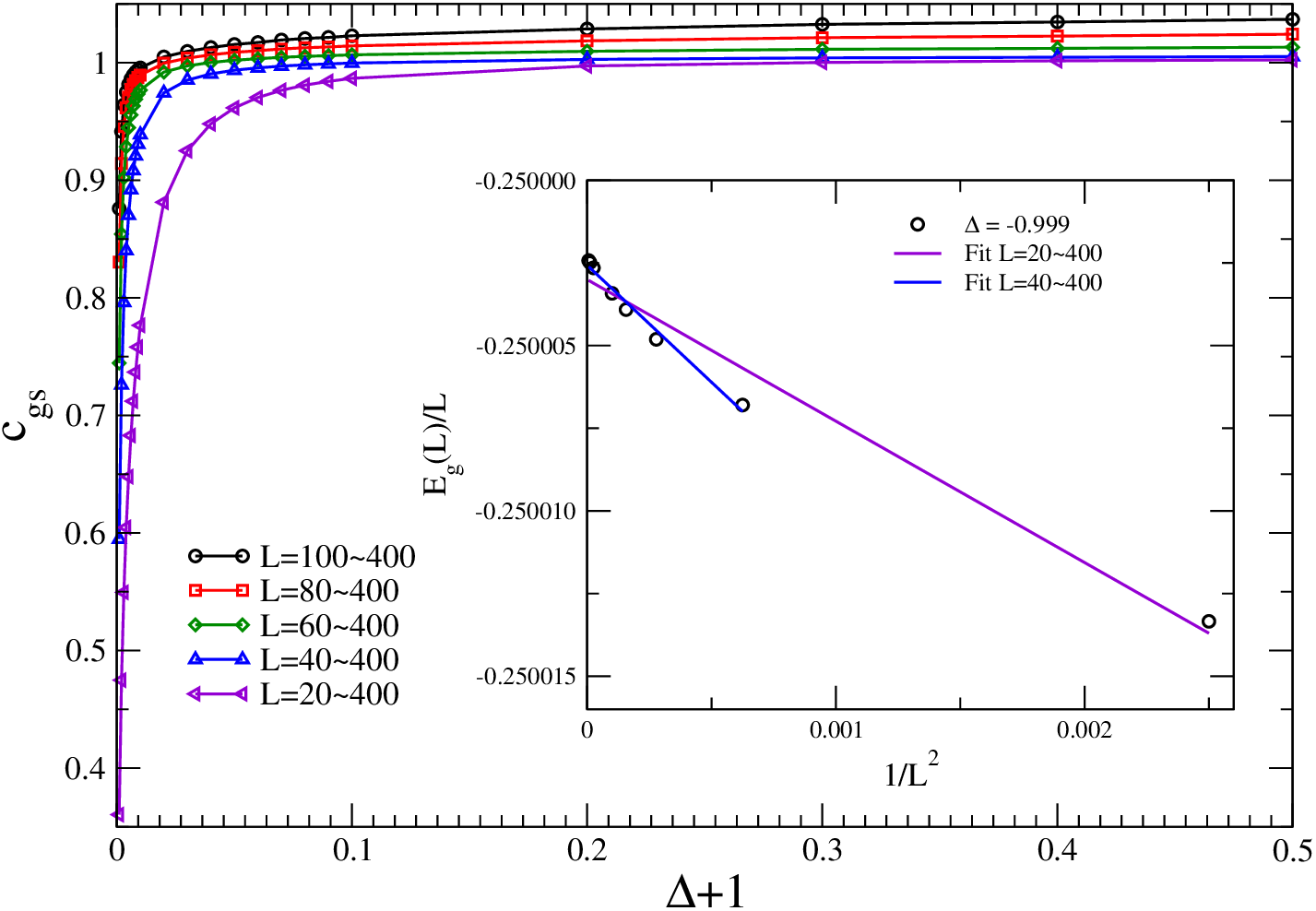}
\caption{(Color online)
$c_{\mathrm{gs}}$ as a function of $\Delta+1$ by fitting Eq.\ref{eq:scaling_d_gs} with data from $L\ge L_c$
and $L_c=20$ (purple left triangle), $40$ (blue right triangle), $60$ (green diamond), $80$ (red square), and $100$ (black circle) respectively.
Inset: Ground state energy  per site for $\Delta=-0.999$ (black circle).
Fitted lines using Eq.\ref{eq:scaling_d_gs} with data from $L=20$ to $400$ (purple line) and $L=40$ to $400$ (blue line) respectively.
}
\label{fig:gs}
\end{center}
\end{figure}

In this section we use the finite-size scaling of the ground state energy to estimate the central charge.
It is well known that when the critical system is described by a CFT with central charge $c$ in the continuum limit,
the ground state energy of a finite system of length $L$ scales as
\begin{equation}
  \frac{E_g(L)}{L} = \epsilon_\infty -\frac{\pi v}{6L^2} c,
  \label{eq:scaling_d_gs}
\end{equation}
where $\epsilon_\infty$ is the ground state energy per site in the thermodynamics limit and $v$ is the spin-wave velocity \cite{Cardy:1986}.
In this work we use ED (for $L\le 20$) and finite-size DMRG (for $L>20$) to obtain the ground state energy $E_g(L)$ of size $L$.
In order to obtain $c_{\mathrm{gs}}$ by fitting $E_g(L)$ with Eq.~\ref{eq:scaling_d_gs}, it is better
to have an independent estimation of the spin-wave velocity $v$. Since we are mainly interested in the influence
of the ferromagnetic point, we first analyze our data using the exact spin-wave velocity,
\begin{equation}
  v_{th}(\Delta)=\frac{\pi\sin(\mu)}{2\mu}.
  \label{v-th}
\end{equation}
which is obtained from the Bethe Ansatz \cite{Alcaraz:1989dt}.  Here $\mu$ is defined via $\Delta = \cos(\mu)$
with $ 0 < \mu < \pi$.
The scenario in which the spin-wave velocity is obtained numerically will be discussed later.
In the inset of Fig.~\ref{fig:gs} we show the fitted lines for $\Delta=-0.999$ using data from $L=20$ to $400$ (purple line)
and $L=40$ to $400$ (blue line) respectively. 
It is clear from the figure that the higher order corrections to Eq.~\ref{eq:scaling_d_gs} are large.
Furthermore, we observe that for $\Delta \in (-1,-0.5]$ all the data behave in a similar fashion.
These higher order corrections can be suppressed by removing smaller sizes data in the fitting procedure.
In Fig.~\ref{fig:gs} we show the fitted $c_{\mathrm{gs}}$ using data with $L\ge L_c$ with $L_c=20,40,60,80$, and $100$ respectively.
We find that when the system is far away from the ferromagnetic point, one has $c_{\mathrm{gs}} \approx 1$ regardless the $L_c$ used.
When approaching the ferromagnetic point, fitted $c_{\mathrm{gs}}$ begins to deviate from 1 and monotonically decreases.
However, larger $L_c$ allows one to reach closer to the ferromagnetic point while maintaining $c_{\mathrm{gs}} \approx 1$.
This behavior is consistent with the observation in the preceding section that there is a $\Delta$ dependent length scale,
which grows larger as one approaches the ferromagnetic point. The system only behaves like one with the XXZ ground state
when the system size is larger than this length scale. This is why larger and larger $L_c$ is needed to obtain the
proper central charge as $\Delta+1 \rightarrow 0^+$. However it is unclear to us how to define a proper $L_c$ from the
$\xi^F_c(\Delta)$ obtained in the preceding section.

In order to ensure that the behavior above is due to the ferromagnetic point and not due to the inaccuracy of the data,
it is important to have some independent check on the quality of the data. We have performed the following two tests.
First, we investigate how our results depend on the $m$ used. For finite-size DMRG calculation,
it is expected that the energy is very accurate once $m$ is large enough.
In our calculation we ensure that $m$ is large enough so that the fitted values of $c_{\mathrm{gs}}$ and $d_{\mathrm{gs}}$ are not sensitive to $m$.
Second, we compare the fitted $\epsilon_\infty^{\mathrm{fit}}$ to the exact energy per site from the Bethe Ansatz \cite{Alcaraz:1989dt}:
\begin{equation}
  \epsilon_\infty(\Delta)=\frac{\cos(\mu)}{4}-\frac{\sin(\mu)}{\mu}
  \int_{-\infty}^{\infty} \frac{\mu\sin(\mu)dx}{2\cosh(\pi x)[\cosh(2\mu x)-\cos(\mu)]},
\end{equation}
where $\mu$ was defined below equation (\ref{v-th}).  We find that the absolute error $|\epsilon_\infty^{\mathrm{fit}}(\Delta)-\epsilon_\infty(\Delta)|$
is at most at the order of $10^{-6}$.

\section{Spin-wave velocity, excited states energies, and scaling dimension}
\label{sec:scaling}

\begin{figure}[btp]
\begin{center}
\includegraphics[width=0.9\columnwidth]{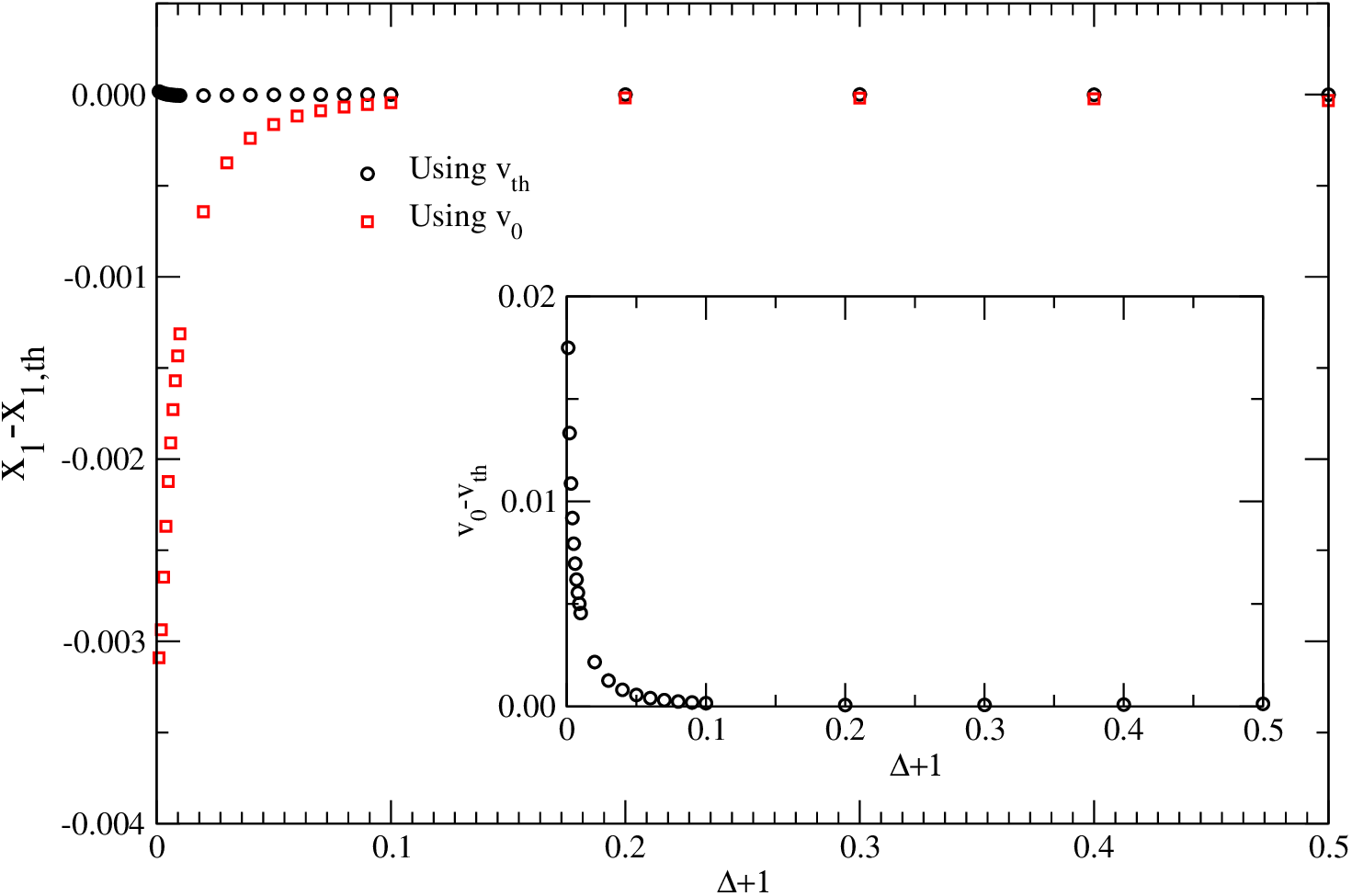}
\caption{(Color online)
Absolute error of the scaling dimension $x_1-x_{1,th}$ obtained by using $v_{th}$ (black circle) and $v_0$ (red square) respectively.
Inset: Absolute error of the spin wave velocity $v_0-v_{th}$.
}
\label{fig:scaling}
\end{center}
\end{figure}

In the analysis in the preceding section, we used the exact spin-wave velocity for the fitting of the central charge.
In general, however, the spin-wave velocity can only be estimated numerically.
In the conventional approach one first obtains an $L$-dependent velocity from the ground state energy
and the lowest energy with momentum $k=2\pi/L$
\begin{equation}
  v(L)=\frac{L}{2\pi}\left[ E\left(k=\frac{2\pi}{L}\right)-E(k=0) \right],
\end{equation}
then one uses the following scaling ansatz
\begin{equation}
  v(L)=v+a\frac{1}{L^2}+b\frac{1}{L^4}
\end{equation}
to obtain an extrapolated value $v_0$ of the spin-wave velocity in the thermodynamic limit\cite{Hijii:2005co}. 
While it is difficult to keep
momentum quantum number in conventional DMRG, we find that for the XXZ model the lowest energy
state with $k=2\pi/L$ is always the first excited state in the $S^{tot}_z=0$ sector, which can be calculated reliably with DMRG.
In the inset of Fig.~\ref{fig:scaling} we plot the absolute error $v_0-v_{th}$ of the spin-wave velocity.
We find that the error is very small for $\Delta+1 \ge 0.1$ but grows rapidly as one approaches the ferromagnetic point.
This is understandable since the spin-wave velocity should approach zero at the ferromagnetic boundary
which makes it very difficult to be accurately determined by the numerical simulation.

From the results in the preceding three sections we find that the value of central charge
extracted from typical simulation might deviate from the CFT prediction when the system
is close to the ferromagnetic point. This might make it more difficult to identify the
underlying CFT from the value of central charge alone.
Another quantity which can be used to identify the CFT is the scaling dimension
of the primary field. It is known that the excited state energies $E_n(L)$ are related to
the scaling dimension of a certain primary field of the CFT via
\begin{equation}
  E_n(L)-E_g(L)=\frac{2\pi v}{L} (x_n+m+m^\prime),
  \label{eq:scaling_dimension}
\end{equation}
where $x_n$ is the scaling dimension and $m$ and $m^\prime$ are integers.
For the XXZ model the smallest scaling dimension is
\begin{equation}
  x_{1,th}=\frac{\pi-\mu}{2\pi} .
\end{equation}
 The corresponding excited state is the lowest energy state
in the $S^{tot}_z=1$ sector and has momentum $k=\pi$. By using finite-size DMRG
with $U(1)$ symmetry the energy $E_1(L)$ can be evaluated accurately.
To remove the sub-leading correction in Eq.\ref{eq:scaling_dimension} we first obtain
a $L$-dependent scaling dimension defined as
\begin{equation}
  x_n(L)= \frac{ E_1(L)-E_g(L)}{2\pi v},
\end{equation}
and then extrapolate to the thermodynamic limit using
\begin{equation}
  x_n(L)=x_n+a\frac{1}{L^2}+b\frac{1}{L^4}.
\end{equation}

In Fig.~\ref{fig:scaling} we plot the absolute errors $x_1-x_{1,th}$
of the results obtained by using exact velocity (black dot) and estimated velocity (red dot) respectively.
We find that the error is extremely small when the exact velocity is used in the whole parameter regime
including the ferromagnetic point. This suggests that Eq.~\ref{eq:scaling_dimension}, the finite-size
scaling formula of the scaling dimension, is less influenced by the ferromagnetic point provided that
the exact spin-wave velocity is used. However, as shown above, the ferromagnetic point does influence
the estimation of the spin-wave velocity. As shown in Fig.~\ref{fig:scaling} this makes the estimated
scaling dimension less accurate near the ferromagnetic point.

\section{Summary and discussion}
\label{sec:sum}

In summary we investigate how the numerical estimation of the central charge $c$ is influenced
by the non-trivial logarithmic scaling of the entanglement entropy of the nearby ferromagnetic point.
From entanglement point of the view, the nontrivial scaling at the ferromagnetic point gives rise to
a jump of the central charge from $c=1$ predicted by CFT to $c^F_{\mathrm{eff}}=1.5$ of the ferromagnetic point.
In particular we use the entanglement entropy scaling of a finite and an infinite system, as well as the
finite-size scaling of the ground state energy to estimate $c$. We find that all methods are influenced by
the nearby ferromagnetic point but different crossover behavior appears. We also find that
the nontrivial scaling of the ferromagnetic point only manifests itself in the first two methods,
 which are entanglement based.
In the following we briefly summarize the crossover behavior and the proper procedure to estimate $c$ within each method.
We also suggest that one should employ all three methods and use the consistency between different methods
as an additional check for the accuracy of the value of $c$.

In any finite-size calculation we observe a smooth crossover between these two values.
We show that the proper procedure is that for a fixed $L$ one need to increase $m$ until a saturated value of $c$ is reached.
One then increases $L$ until $c$ is not sensitive to the changes in $L$. If the value of $c$ is still changing when one
reaches the largest $L$ one can calculate, the corresponding $c$ value cannot be trusted.
When iDMRG is used, however, a different crossover behavior appears. Here we show that there is a $\Delta$ dependent
crossover length scale $\xi^F_c(\Delta)$ such that for scales above the system scales according to the CFT prediction while
for scales below the system scales according to the ferromagnetic point. For any model with a nearby
ferromagnetic like point we suggest that one should increase $m$ (hence $\xi$)
gradually to check if such a crossover behavior appears or not.

We also show that the conventional method to use finite-size scaling of the ground state energy to determine $c$
also suffers from the influence of the nearby ferromagnetic point. It induces a sub-leading correction which seems to
have an opposite sign with respect to the logarithmic correction due to the marginally irrelevant operator.
 For the unknown model,
the sign of the correction can be used to detect the existence of such an influence. Here we suggest that one should just
discard smaller size data.
In addition, we show that the finite-size scaling relation between the excited state energy and the scaling dimension
of the primary field is less influenced by the ferromagnetic point, provided that exact spin-wave velocity is used.
The numerical estimation of the velocity itself, however, is also influenced by the ferromagnetic point.

\section*{Acknowledgement}
We thank Andreas M. La\"uchli for fruitful discussions.
We acknowledge the support from National Center for Theoretical Science (NCTS) 
and National Science Council (NSC), Taiwan. 
Ian McCulloch acknowledges support from the Australian Research Council Discovery Projects funding scheme (project number DP1092513).

\section*{References}

\end{document}